\documentclass[11pt]{article}
    \makeatletter
    \renewcommand\section{\@startsection {section}{1}{\z@}%
                                       {-3.5ex \@plus -1ex \@minus -.2ex}%
                                       {2.3ex \@plus.2ex}%
                                       {\normalfont\fontfamily{phv}\fontsize{16}{19}\bfseries}}
    \renewcommand\subsection{\@startsection{subsection}{2}{\z@}%
                                         {-3.25ex\@plus -1ex \@minus -.2ex}%
                                         {1.5ex \@plus .2ex}%
                                         {\normalfont\fontfamily{phv}\fontsize{14}{17}\bfseries}}
    \renewcommand\subsubsection{\@startsection{subsubsection}{3}{\z@}%
                                        {-3.25ex\@plus -1ex \@minus -.2ex}%
                                         {1.5ex \@plus .2ex}%
                                         {\normalfont\normalsize\fontfamily{phv}\fontsize{14}{17}\selectfont}}
    \makeatother

    \usepackage{amsmath}
    \DeclareMathOperator*{\argmax}{argmax}
    \usepackage{amsthm}
	\usepackage{graphicx}
	\usepackage{enumerate}
	\usepackage{xcolor}
	\usepackage{natbib} 
	\usepackage{url} 
    \usepackage[symbol]{footmisc}

	\usepackage{listings}
    \usepackage{color}
    
    \definecolor{dkgreen}{rgb}{0,0.6,0}
    \definecolor{gray}{rgb}{0.5,0.5,0.5}
    \definecolor{mauve}{rgb}{0.58,0,0.82}
    
    \usepackage{hyperref}

\begin{document}
		
			\def\spacingset#1{\renewcommand{\baselinestretch}
			{#1}\small\normalsize} \spacingset{1}

\title{\textbf{Transfer Learning for Individualized Treatment Rules: Application to Sepsis Patients Data from eICU-CRD and MIMIC-III Databases}}

\author{Andong Wang$^1$\thanks{Corresponding Author. Email address: andong@unc.edu}, Kelly Wentzlof$^2$\thanks{Equal contribution.}, Johnny Rajala$^{3{\dagger}}$, Miontranese Green$^4$, \\
Yunshu Zhang$^1$, and Shu Yang$^1$\\
\\
$^1$North Carolina State University\\
$^2$Indiana University Bloomington\\
$^3$University of Maryland, College Park\\
$^4$California State University, Long Beach}
\date{}
\maketitle

\begin{abstract}

Modern precision medicine aims to utilize real-world data to provide the best treatment for an individual patient. An individualized treatment rule (ITR) maps each patient's characteristics to a recommended treatment scheme that maximizes the expected outcome of the patient. A challenge precision medicine faces is population heterogeneity, as studies on treatment effects are often conducted on source populations that differ from the populations of interest in terms of the distribution of patient characteristics. Our research goal is to explore a transfer learning algorithm that aims to address the population heterogeneity problem and obtain targeted, optimal, and interpretable ITRs. The algorithm incorporates a calibrated augmented inverse probability weighting (CAIPW) estimator for the average treatment effect (ATE) and employs value function maximization for the target population using Genetic Algorithm (GA) to produce our desired ITR. To demonstrate its practical utility, we apply this transfer learning algorithm to two large medical databases, Electronic Intensive Care Unit Collaborative Research Database (eICU-CRD) and Medical Information Mart for Intensive Care III (MIMIC-III). We first identify the important covariates, treatment options, and outcomes of interest based on the two databases, and then estimate the optimal linear ITRs for patients with sepsis. Our research introduces and applies new techniques for data fusion to obtain data-driven ITRs that cater to patients' individual medical needs in a population of interest. By emphasizing generalizability and personalized decision-making, this methodology extends its potential application beyond medicine to fields such as marketing, technology, social sciences, and education.
 
\end{abstract}
			
\noindent%

\spacingset{1.5} % DON'T change the spacing!

\newpage

\section{Introduction} \label{s:intr}
Under FDA's 21st-Century Cures Act, the field of precision medicine is developed to utilize real-world data to provide evidence based and data-driven optimal treatment for each individual patient \citep{Kosorok2019}. However, a ongoing challenge facing modern medicine is that studies on outcomes of potential treatments are conducted for a source population that is different from some target population for which the treatment will be implemented in terms of the distribution of patients' demographics, characteristics, and health factors. While treatment may be proven effective after testing on a sample population, it may not have the same optimal effect for all patients within other populations. This problem is often referred to as population heterogeneity, or covariate shift and it is widely studied in social sciences and biological sciences (e.g. \citet{Nagin2000}; \citet{Ryall2012}). In medicine, a typical example of population heterogeneity consists of an experimental clinical trial population and a real-world population. Despite experimental data having strong internal validity, it has limited external validity and introduces bias when applied to other populations due to its limited sample pool from inclusion and exclusion criteria \citep{Rothwell2005}. As opposed to experimental data, using real-world data and observational data provides a sample that is representative of the larger population. Thus, the goal of this research is to use knowledge learned about one population to inform decisions for another, a technique referred to as transfer learning. We will investigate transfer learning applied to two large data sources, eICU-CRD and MIMIC-III, to learn targeted, optimal, and interpretable individualized treatment rules (ITRs) for patients with sepsis. We hope to inspire other researchers to apply our framework to populations of patients with other diseases or adapt it to other fields of study. 

To achieve our research goals, we study the transfer learning framework. Based on potential outcome framework in causal inference \citep{Imbens2015}, transfer learning combine two key methodologies: the Augmented Inverse Probability Weighting (AIPW) and calibration weighting (CW). The AIPW estimator estimates the average treatment effect (ATE), which is the average expected outcome for a patient population under one specific treatment rule. The AIPW estimator has the desirable property of double robustness: it provides a consistent estimation even if either the outcome regression model or the propensity score model is misspecified \citep{Glynn2010}. CW is an intuitive method that corrects the selection bias of the source population -- hence addressing the population heterogeneity -- by calibrating the covariates of patients in the source population to assimilate the target population’s patient covariates distributions using entropy balancing \citep{Hainmueller2012}. 

To demonstrate the power of the framework in a real-world problem context, we apply transfer learning to two populations of patients with sepsis. Sepsis is a life-threatening complication of an infection, leading to 270,000 deaths each year in America. Furthermore, there is approximately one patient suffering from sepsis in every three hospital deaths. Providing precise treatment to sepsis patients is detrimental to preventing further medical complications such as organ failure and possible death. 

For our application on producing the best ITR for sepsis patients, we introduce the MIMIC-III and eICU-CRD databases that contain sepsis patient data from two populations. We identify the source and target populations, patient covariates to be included in the model, treatment options, and the outcome of interest. Then, we produce a target, optimal, interpretable ITR. 

We also conduct a simulation study in which we consider a population with only two covariates: height and age. We demonstrate calibration weight computation and integration of these calibrations weights and the AIPW estimator in a value function. We run the genetic algorithm to maximize the value function to obtain the maximizer, i.e. the optimal ITR for the target population. Then, we assess the performance of our method.

Finally, we summarize the effectiveness of the transfer learning framework and share the implications of our findings. 

The rest of the paper is organized as follows: In Section \ref{s:fram}, we define notations and outline the theoretical assumptions as the premise of our research. We then compare estimators for the ATE. In Section \ref{s:mthd}, we delve into the three main components of the methodology: CW, Value Function, and Genetic Algorithm. In Section \ref{s:appl}, we apply the framework to real-world data of two populations of patients with sepsis and compile the application results. In Section \ref{s:sims}, we validate the utility of our framework by conducting a simple simulation study. In Section \ref{s:disc}, we discuss the advantages and limitations of our framework. 

All relevant R code for the medical application and the simulation study are provided in Supplementary Material.

\section{Statistical Framework} \label{s:fram}

To study the average effect of a treatment rule and eventually optimize it, we adopt some basic concepts from the potential outcomes framework for causal inference in \citep{Imbens2015} as building blocks for our research. The ATE measures the difference between the average outcome that would be achieved if all individuals in the population were to receive treatment and if all were to receive control, given that the treatment is binary. 

We start with an overview of the notation that will be used throughout this paper. Then, we will move to the assumptions that must hold true for us to proceed to develop our methodology.

\subsection{Notation} \label{ss:nota}

Suppose we have a general population of $N$ patients. Our population of interest within the general population has $n\leq N$ patients. Each patient is indexed by $i$ and $I_n$ represents the indices set of patients in the population of intereset. The baseline covariates of these patients are specified in matrix $X,$ with $X_i$ being the vector of covariates for patient $i.$ We refer to covariates as pre-treatment attributes or features of each patient such as age, gender, medical history, etc. Covariates help explain variation in outcomes making estimates more precise and allowing researchers to separate the units into subgroups detecting a difference between the two treatment groups. The treatment assignment vector $A \subset \{0,1\}^{n}$ denotes what intervention each patient gets. We will be considering binary treatment, with $A_i = 1$ if the patient receives treatment, and $A_i=0$ if they receive control. $Y$ represents patient outcome, i.e. the greater the $Y_i$, the better the patient reacts to the treatment assigned to them. 

Following the potential outcome framework proposed in \citet{Imbens2015}, the individual treatment effect is defined as the difference in the outcome if the patient is given treatment, $Y^*(1)$, over the outcome if the same patient is given control, $Y^*(0)$. For each patient, one of these two outcomes can be observed, referred to as the "observed outcome"; the other outcome will be missing, referred to as the "missing outcome" or the "counterfactual outcome". For patient $i$, the potential outcome under treatment $a$ is represented as $Y^*(A_i=a)$. Hence, we can represent the ATE by taking the expectation of differences between potential outcomes: $E[Y^*(1)-Y^*(0)]$.

When estimating the ATE, $\tau$, we consider the propensity score. The propensity score is the probability unit $i$ receives active treatment given its covariates, $\pi(X_i)$. In randomized clinical trials, researchers control the probability of each patient receiving treatment, which is typically fixed across patients, regardless of patient characteristics; in an observational study, a patient's propensity score may vary depending on their characteristics, introducing assignment bias. In Section \ref{ss:esti}, we discuss a variety of estimators for the ATE, including the naive estimator, the inverse probability weighting (IPW) estimator, the outcome regression (OR) estimator, and the augmented inverse probability weighting (AIPW) estimator, then we explore how these estimators address the assignment bias.

We denote the population to which the patient belongs with the binary indicator $S$ . If patient $i$ is from the source population then $S_i = 0$; if the patient is from the target population then $S_i = 1$. When population heterogeneity exists, covariate distribution differs between populations, i.e. $Pr(X=x|S=1)\neq Pr(X=x|S=0)$.  This is also referred to as selection bias or sampling bias. We address selection bias with calibration weighting in Section \ref{ss:calw}.

We use $d$ to denote an ITR and $d_\eta$ to denote a linear ITR with $\eta$ specifying the vector of covariate coefficients that uniquely identifies the rule. 

\subsection{Assumptions} \label{ss:assu}

\newtheorem{assumption}{Assumption}
\begin{assumption}

Stable Unit Treatment Value: $Y = Y^{*}(1)(A) + Y^{*}(0)(1-A)$ 

\end{assumption}

Meeting the Stable Unit Treatment Value Assumption (SUTVA) allows researchers to use multiple units within one study.  SUTVA incorporates two main components. 

The first component of SUTVA is that the treatment of one unit does not affect the outcome of another unit. This is typically ensured when researchers separate the participants of a study, therefore, reducing the likelihood of the participants' effects intermingling. If participants interact resulting in outcomes that are different if the participants had not come in contact with each other, then SUTVA is violated. 

The second component of SUTVA is that the researchers must minimize any differences in the efficacy and the method of administering the treatment. For example, in a drug trial, researchers must ensure that patients in the treatment group all receive the treatment of the same strength. If the efficacy of the treatment varies within this singular treatment group, then SUTVA is violated.  

Note that there can be various levels of the treatment, however, each person in a single given treatment must receive a consistent treatment with the others in the same treatment group. 

\begin{assumption}

No Unmeasured Confounding: $(Y^{*}(1), Y^{*}(0) )\perp\!\!\!\perp\ A \mid X$

\end{assumption}

The No Unmeasured Confounding (Conditional Exchangeability) assumption entails that all the variables that influence both the treatment assignment and the outcome of interest are measured and accessible in our data. If there are unmeasured confounding variables, our estimation of the causal effect will be biased. This assumption is one of the most commonly violated assumptions in causal inference.

\begin{assumption}

Positivity of Treatment Assignment: 0 $<$ $\Pr(A=a \mid X=x)$ $<$ 1 for all $x$ and $a = 0,1$

\end{assumption}

The Positivity assumption requires every patient to have a positive probability of being assigned to treatment or control. If this assumption is violated, all patients could be assigned to one treatment group, rendering inference impossible. This is essential as we model the counterfactual and estimate the treatment effect.

\begin{assumption}

Transportability: $E[Y^*(A=a)\mid S=1, X=x] = E[Y^*(A=a) \mid X=x]$ for all $x$ and $a = 0,1$

\end{assumption}

Transportability describes the ability to "transport" causal effect estimated from a random clinical trial or observation study done on the source population to a target population. It requires the ATE to be consistent across populations. In recent literature, transportability is sometimes referred to as generalizability, external validity, or recoverability. These terms have slightly different definitions concerning the overlap between populations and there have been discussions of the differences as seen in \citet{Colnet2020}.  

\begin{assumption}
Common Support: $Pr(S = 1|X) >0$
\end{assumption}

The common support assumption entails that for the inference to be transportable, the support for the source population covariate distribution is required to overlap the support for the target population covariate distribution. 

Using these assumptions and the following methods for estimation, we hope to find the ITR that maximizes the patient outcome for the target population $\{X_i, Y_i^{*}(0),Y_i^{*}(1)\}_{i=1}^{N}$.

\subsection{Estimators} \label{ss:esti}

Now we will consider the various estimators for the ATE, $\tau$, and how each handles assignment bias that is introduced when the propensity score is not predetermined. 

\subsubsection{Naive Estimator} \label{sss:naiv}

The naive estimator, $\hat{\tau}_{0}$, takes the difference between the average observed outcome of patients in the treatment group and the average observed outcome of patients in the control group to obtain the ATE. Let $n_1$ be the number of patients in the treatment group and $n_0$ be the number of patients in the control group, then the naive estimator is represented as:

\begin{equation}
\hat{\tau}_{0}=\frac{1}{n_1}\sum_{i=1}^{n_1}Y_i - \frac{1}{n_0}\sum_{j=1}^{n_0}Y_j
\end{equation}

\noindent
The naive estimator provides a relatively accurate estimation of the ATE for randomized experiments because there is no assignment bias that needs to be addressed. However, with observational data, the naive estimator is biased and performs poorly because there is no way to consider the covariates that the treatment depends on (i.e., there is no way to minimize the assignment bias). 

\subsubsection{Inverse Probability Weighting Estimator} \label{sss:ipwe}

The Inverse Probability Weighting (IPW) estimator, $\hat{\tau}_\text{IPW}$, estimates ATE with the objective to address the assignment bias and is formulated as:

\begin{equation}
\hat{\tau}_{\text{IPW}}=\frac{1}{n}\sum_{i=1}^{n}\{\frac{A_iY_i}{\hat{\pi}(X_i)}\} 
\end{equation}

\noindent
The IPW estimator weights each patient's treatment effect based on their covariates. The weights, $\frac{1}{\hat{\pi}(X_i)}$, adjusts for the probability of patient $i$ receiving treatment. The more likely a patient is to receive treatment, the higher the propensity score, the less the weight. The IPW estimator is unbiased if we correctly specify the propensity score model.

\subsubsection{Outcome Regression Estimator} \label{sss:oreg}

The Outcome Regression (OR) estimator, $\hat{\tau}_\text{OR}$, estimates ATE by modeling the outcome based on observed values for covariates, as shown in \eqref{oreg}. We denote $m$ as the true mapping from covariates to outcome. However, in reality, it is more likely that $m$ is unknown and needs to be estimated. Thus, we use $\hat{m}(X_i)$ to represent the estimated outcome for patient $i$ based on the OR model and their covariates. The fitted OR model can be parametric (e.g., linear regression, logistic regression) or nonparametric (e.g., machine learning methods). 

\begin{equation} \label{oreg}
\hat{\tau}_{\text{OR}}=\frac{1}{n}\sum_{i=1}^{n}\hat{m}(X_i)
\end{equation}

\noindent
If the model $m$ is correctly specified, then we can better estimate the missing outcomes and the treatment effect. Although the OR estimator is efficient, it tends to be nonrobust due to the misspecification of the model.

\subsubsection{AIPW Estimator}  \label{sss:aipw}

The Augmented Inverse Probability Weighting (AIPW) estimator, $\hat{\tau}_\text{AIPW}$, estimates the ATE by augmenting the IPW estimator with OR, as shown in \eqref{aipw}, achieving the doubly robust property. 

\begin{equation} \label{aipw}
\hat{\tau}_\text{AIPW}=\frac{1}{n}\sum_{i=1}^{n}\{\frac{A_iY_i}{\pi(X_i)}-\frac{A_i-\pi(X_i)}{\pi(X_i)}m(X_i)\}
=
\frac{1}{n}\sum_{i=1}^{n}\{\frac{A_i(Y_i-m(X_i))}{\pi(X_i)}+m(X_i)\}
\end{equation}

\noindent
The double robustness of the AIPW estimator proposed in \citet{Glynn2010} suggests that if either the propensity score model or the outcome regression model is misspecified, the AIPW estimator will remain unbiased. In the equation (4) above, two mathematical representations of the AIPW estimator are shown to demonstrate the doubly robust property. In the expression on the left side, if the propensity score model is correctly specified, then $\pi(X_i)$ closely approximates $A_i$, which will cancel out the outcome regression component, leaving only the correctly specified IPW model. Similarly, in the expression on the right side, if the outcome regression model is correctly specified, then $m(X_i)$ closely approximates $Y_i$, which will cancel out the IPW estimator component, leaving only the correctly specified OR model.

In our application, we employ the AIPW estimator to estimate the ATE for its double robustness.

\section{Proposed Method} \label{s:mthd}

\subsection{Calibration Weighting} \label{ss:calw}

When covariate heterogeneity exists between patients from the source population and the target population, the optimal ITR we find for the source population will likely not be optimal for the target population. In other words, the optimal ITR learned from clinical trials or observation studies possesses internal validity and lacks external validity, which means this ITR will not lead to the best outcome for patients awaiting treatment outside of that clinical trial or observation study. This issue can be referred to as a selection bias or sampling bias since the sampling distribution of the source population differs from that of the target population. 

To correct this sampling bias, we use entropy balancing methods introduced in \citet{Hainmueller2012} to compute calibration weights based on the covariates information from patients of both the source and target population. The entropy balancing weighting is a trusted method for balancing covariates and has been studied and applied recently in \citet{Chu2023} and \citet{Wu2023}. 

After computing the weights, we then assign these weights to individual units from the source population, such that the covariate distributions of the two populations are similar. A weighted, balanced source population gives us a better estimation of the treatment effects for the target population.

\subsection{Value Function} \label{ss:vfxn}

The value function estimates the total treatment effect for the whole population given an ITR, $d_\eta$. It is used to evaluate the quality of an ITR. As shown in \eqref{vfxn}, our value function combines the AIPW estimator and the calibration weights, $\hat{w}$, which addresses both the treatment assignment bias and the sampling bias. 

\begin{equation} \label{vfxn}
\hat{V}(d_{\eta};\hat{w},I_n) =  \sum_{i\epsilon{I_n}}\hat{w_i}([\frac{A_id(X_i;\eta)}{\hat{\pi}(X_i)} + \frac{(1-A_i)\{1-d(X_i;\eta)\}}{1-{\hat{\pi}}(X_i)}][Y_i-\hat{m}(X_i)]+\hat{m}(X_i)) 
\end{equation}

We integrate the two components to construct calibrated augmented inverse probability weighting (CAIPW) estimator which estimates the average weighted outcome of patients given treatments based on a specific ITR, $d_\eta$. Note that similar to AIPW estimator, CAIPW estimator consists of estimate of the propensity score $\hat{\pi}(X_i)$ and outcome estimate from outcome regression $\hat{m}(X_i)$, which can be obtained through logistic regression and linear regression respectively. Additionally, CAIPW weights are normalized, so the summation of the weighted treatment effect in the value function yields the ATE.

\subsection{Genetic Algorithm} \label{ss:galg}

The value function has a greater value for an ITR that performs better in optimizing patient outcome. Our goal is to find the ITR that maximizes the value function. 

\begin{equation} \label{dopt}
d^\text{opt} = {\argmax_{d_\eta}}  \hat{V}(d_{\eta};\hat{w_i},I_n)
\end{equation}

In order to obtain the optimal ITR in \eqref{dopt}, $d^\text{opt}$, we use Genetic Algorithm (GA) to solve the value function maximization problem, specifically, using the "rgenoud" R package developed by \citet{Mebane2011}. 

GA is a search-based optimization tool inspired by the mechanism of biological evolution and natural selection with a wide range of applications across disciplines as discussed in \citet{Katoch2021}. It is a population-based algorithm, which means it maintains a population of candidate solutions, allowing for both the search for a new solution space and the refinement of outstanding solutions in the current solution space. This makes the search process more robust, and overcoming the local maxima easier. GA is flexible with respect to the type of objective function it can optimize. It is a good choice for our complex and noisy value function and can generate high-quality solutions.

\section{Medical Application}  \label{s:appl}

In this section, we apply our proposed method to two large-scale public databases to demonstrates its practical relevance in solving real-world problems.  

\subsection{Databases} \label{ss:dtbs}

Medical Information Mart for Intensive Care (MIMIC-III) is a freely accessible single-center database consisting of desensitized health record data of over 40,000 patients who stayed in the intensive care units of the Beth Israel Deaconess Medical Center from 2001 to 2012. 

The electronic Intensive Care Unit Collaborative Research Database (eICU-CRD) is a freely accessible multi-center database for critical care research. It consists of desensitized health record data of over 200,000 patients who were treated in intensive care units all over the U.S. under the Phillips eICU program from 2014 to 2015.

A randomized clinical trial conducted in a single center tends to follow a simpler design compared to a multi-center trial, but it often has limited external validity. Since our methodology aims to expand external validity, in our application, we use the single-center data from MIMIC-III as our data for the source population and the multi-center data from eICU-CRD as our data for the target population.

\subsection{Data Pre-processing} \label{ss:dtpp}

After an Exploratory Data Analysis of the raw data, we removed duplicate observations of patients in both datasets to keep only the baseline entries of each patient before any treatment is applied. This leaves us with 20,955 unique patients' data in MIMIC-III and 21,995 unique patients' data in eICU-CRD.

We then reformatted and rescaled certain variables for consistency across the two datasets and removed all non-binary categorical variables due to their incompatibility with our methods. We also removed variables with over 60\% missing entries and imputed the rest of the missing data with the MICE (Multivariate Imputation by Chained Equations) algorithm, introduced in \citet{Azur2011}. The MICE algorithm fills in the incomplete variable based on observed variables multiple times in an iterative manner. Upon visually inspecting the covariate lists of both datasets, we kept the covariates that are shared by both datasets due to the requirement of calibration weighting. This leaves us with usable data from thirteen common variables in both datasets: re-admission (binary), age, weight at admission, temperature in Celsius, mean blood pressure, respiratory rate, sodium, glucose, blood urea nitrogen, creatinine, bilirubin, albumin, and white blood cell count. Two other variables, mechanical ventilation (binary) and death (binary) are only available in the MIMIC-III dataset but are kept as treatment and outcome variables in the source data. For more detailed reasoning for these decisions, see Section \ref{ss:tmoc}.

\subsection{Treatment and Outcome} \label{ss:tmoc}

Out of the 15 variables available after data pre-processing, we chose mechanical ventilation as treatment and death as outcome. 

The interplay between mechanical ventilation and sepsis has been discussed in recent medical research. Sepsis is responsible for approximately 70\% of acute respiratory distress syndromes (ARDS). \citet{Zampieri2017} suggests that while a mechanical ventilator is necessary to support breathing for patients with sepsis-induced ARDS, suboptimal use of mechanical ventilators can cause lung injuries, which further contributes to a downward spiral of sepsis-related organ failures. The use of mechanical ventilation should be tailored to the patient's individual characteristics and conditions to improve. The emphasis on personalization makes mechanical ventilation a treatment of interest for our application. 

The choice of patient death as the outcome variable in our application was an intuitive one. However, there are three caveats with this outcome variable:

1. Information on patient deaths is only available in MIMIC-III, our source population, which means we can only evaluate the effect of our optimal ITRs using estimated outcomes in our target population. 

2. Death is coded in 3 scenarios in MIMIC-III: 1) Died in hospital; 2) Died within 48 Hours after discharge; 3) Mortality within 90 Days. We decided to integrate the 3 scenarios into a binary outcome variable that classifies a patient as dead (death = 1) if they fit at least one of the scenarios and not dead (death = 0) otherwise. 

3. Since we are maximizing the value function to yield the best outcome for patients, we need to ensure that the outcome variable is greater in value when representing a "better" outcome. In our application, we use (1 - death) as the outcome variable, representing patient survival. Therefore, the outcome variable is 1 if the patient survives, 0 otherwise.

\subsection{Covariates} \label{ss:covr}

The remaining 13 variables (hospital re-admission, mean blood pressure, body temperature, respiratory rate, sodium, glucose, blood urea nitrogen, creatinine, bilirubin, albumin, white blood cell count, weight, and age) are used as patient covariates, serving as the active components of our calibration weighting scheme. 

\begin{figure}[h]
    \centering
    \includegraphics[scale=0.7]{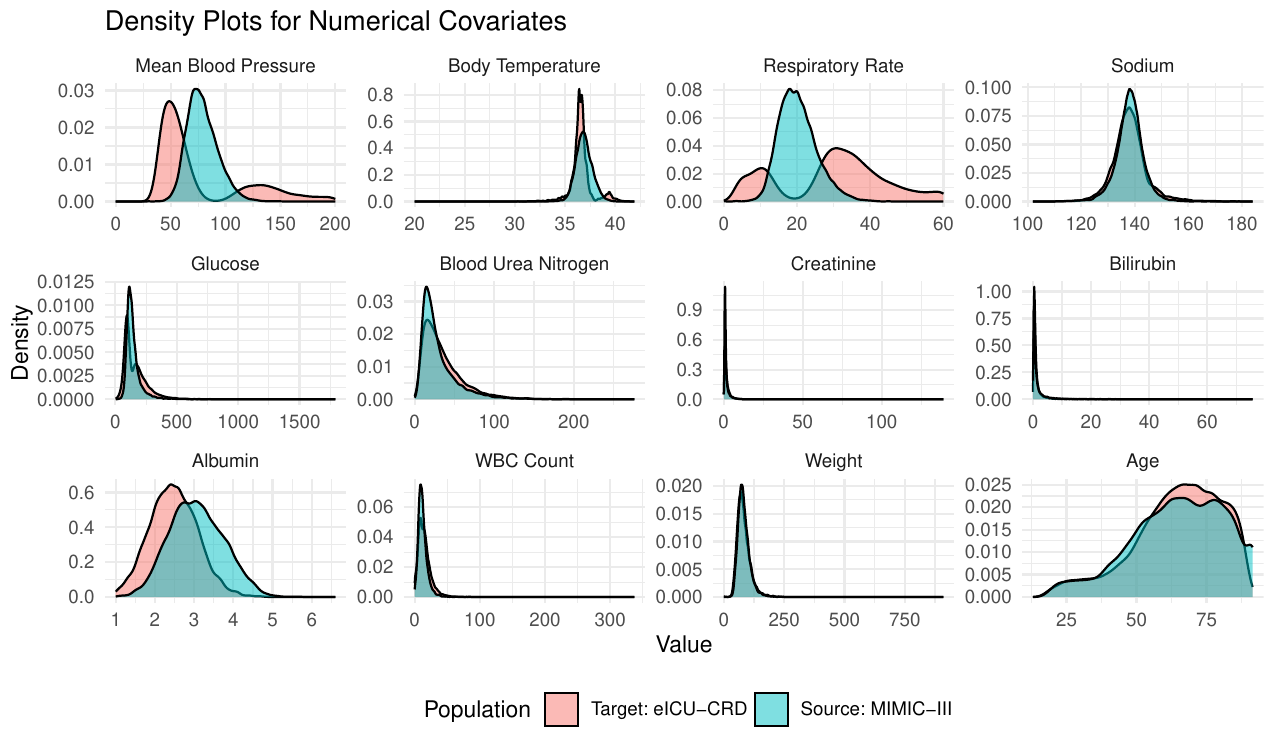}
    \caption{Density Plots demonstrate covariate shifts between the two populations, i.e. population heterogeneity}
    \label{fig:Figure1}
\end{figure}

In Figure \ref{fig:Figure1}, we compare the density curves for each of the 12 numeric variables of the 13 covariates between the target population based on eICU-CRD database and the source population based on MIMIC-III database. We notice that mean blood pressure, body temperature, respiratory rate, glucose, albumin, and age have clear distinctions between distributions for the two populations. The one categorical variable, hospital re-admission, has also shown discrepancy between the two populations. In the source population, approximately 32.77\% of patients had been re-admitted, whereas in the target population, only approximately 4.75\% of patients had been re-admitted. The differences in distribution of these covariates indicate a population heterogeneity problem and calibration weighting is needed to address such problem. 

\subsection{Application Results} \label{ss:rslt}

\subsubsection{Optimal ITR and Survival Rate} \label{sss:opti}

Following our proposed methods, we obtained the entropy balancing calibration weights with covariates, treatment, and outcome data from MIMIC-III and only covariates data from eICU-CRD. Then, we repeatedly ran GA that aims to maximize the value function: 1000 iterations with calibration weights and 1000 iterations with equal weights. Since we do not have data on treatment and outcome for the target population, we evaluate the optimal ITR produced by GA by computing its treatment effect with the value function.

With calibration weighting, the optimal ITR has a value of approximately 0.7616,  while the unweighted optimal ITR has a value of approximately 0.7223. In the context of this application, our results imply that assigning mechanical ventilation treatments to patients using the optimal weighted ITR has led to an approximately 3.93\% increase in survival rate compared to using the optimal ITR without calibration weighting. 

The resulting optimal weighted ITR is illustrated in \eqref{oitr}. After substituting a patient's baseline reading of the covariates into this ITR, if this inequality is true, then such a patient has a positive treatment effect and hence should be given the treatment of interest, and vice versa.

\begin{equation} \label{oitr}
\begin{split}
    0 &< -0.3933\cdot Glucose + 0.6507\cdot Blood Urea Nitrogen + 0.6282\cdot Age  -0.2484\cdot Weight \\ 
    &+ 0.4333\cdot Mean Blood Pressure -0.4738\cdot WBC Count + 0.8800\cdot Respiratory Rate \\ 
    &+ 0.8830\cdot Bilirubin  -0.6220\cdot Sodium  -0.0565\cdot Creatinine -0.7644\cdot ReAdmission \\ 
    &+ 0.5545\cdot Body Temperature + 0.3633\cdot Albumin + 0.1634
\end{split}
\end{equation}

\subsubsection{Covariate Importance} \label{sss:covi}

With the optimal weighted ITR, we can also make rudimentary analyses on which patient covariate is more important to the ITR by comparing the magnitudes of the coefficients. 

To account for the differences in the spreads of covariates' distributions, we adjusted the coefficients by multiplying the standard deviation of the corresponding covariate. 

\begin{figure}[h]
    \centering
    \includegraphics[scale=0.7]{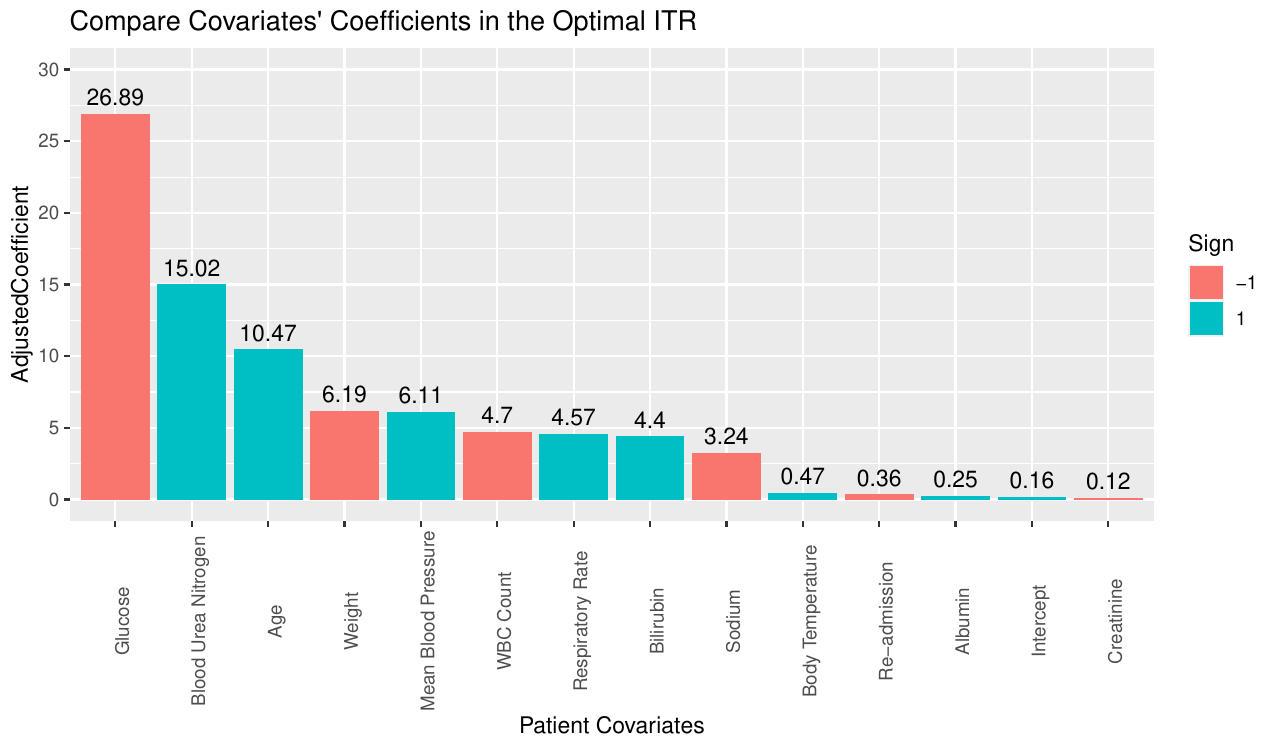}
    \caption{From left to right, the covariates are ranked by the magnitude of their corresponding coefficients in the optimal weighted ITR. Colored by the signs of the coefficients.}
    \label{fig:Figure2}
\end{figure}

Observing Figure \ref{fig:Figure2}, we found that Glucose is the primary covariate in our optimal ITR, and it has a negative coefficient, which indicates that the lower the patient's blood glucose reading, the more likely they will have a positive treatment effect and get assigned treatment. Blood Urea Nitrogen (BUN) is the secondary covariate in our optimal ITR, and it has a positive coefficient, which indicates that the higher the patient's BUN reading, the more likely they will have a positive treatment effect and get assigned treatment. 

\section{Simulation Study}  \label{s:sims}

To further demonstrate the performance of our methods under controlled conditions, we conducted a simulation study. While our medical application showcases the methods' utility in a real-world setting, the simulation study allows us to validate its functionality in setting where covariate distributions and treatment effects are known and can be manipulated. Using generated data for covariates, treatment assignment, and outcome for both populations, we examine how well the CAIPW estimator estimates treatment effects as well as the maximization of the value function to obtain optimal ITRs.

\subsection{Simulation Data} \label{ss:simd}

We generate a dataset with 50,000 rows with each row representing an individual patient. Each patient has height and age as covariates, and both covariates are uniformly distributed. This is our general population. 

We randomly sample 10,000 observations from the general population to use as Real World Data (RWD), i.e. the target population. While usually not available in real practice, we also simulated the target population's treatment and outcome data. 

For treatment data, we first generate propensity scores for all patients in the target population based on a linear model of their covariates. With the propensity scores, we can then model individual Bernoulli trials, to attach binary treatment assignments to the patients.

For outcome data, although in reality, we can only observe one outcome for each patient -- under either treatment or control -- we simulated both outcomes for each patient. In this simulation, we establish a positive treatment effect condition that if a patient is taller than 55 inches and less than 41 years old, they will have a positive response to treatment; otherwise, the treatment will have a negative effect on them, and they should be assigned control. We first start with the control outcome, which is a linear combination of the patient covariates with the addition of a normal error term. Then, we formulate a contrast function that produces a positive value if a patient satisfies the positive treatment effect condition, and vice versa. Lastly, to generate the treatment outcome, we simply add the value of the contrast function to the control outcome. Since we have the treatment assignment and both outcomes, we can identify the observed outcome and the missing outcome for each patient. 

We then generate sampling scores for all observations in the general population minus the target population based on a linear model of their covariates with a bias to observations with greater age value and greater height value. With sampling scores, we can then model individual Bernoulli trials, to decide whether each of these observations will be sampled in a Random Clinical Trial (RCT), i.e. the source population. Because of the bias in the sampling scores, patients sampled in the source population are more likely to be old and tall. Following an actual random clinical trial design, we give each patient in the source population an equal chance to be put into the treatment group (50\%) or the control group (50\%). The outcome data for the source population is generated the same way as for the target population.

\subsection{Simulation Results} \label{ss:simr}

\begin{figure}[h!]
    \centering
    \includegraphics[scale = 0.7]{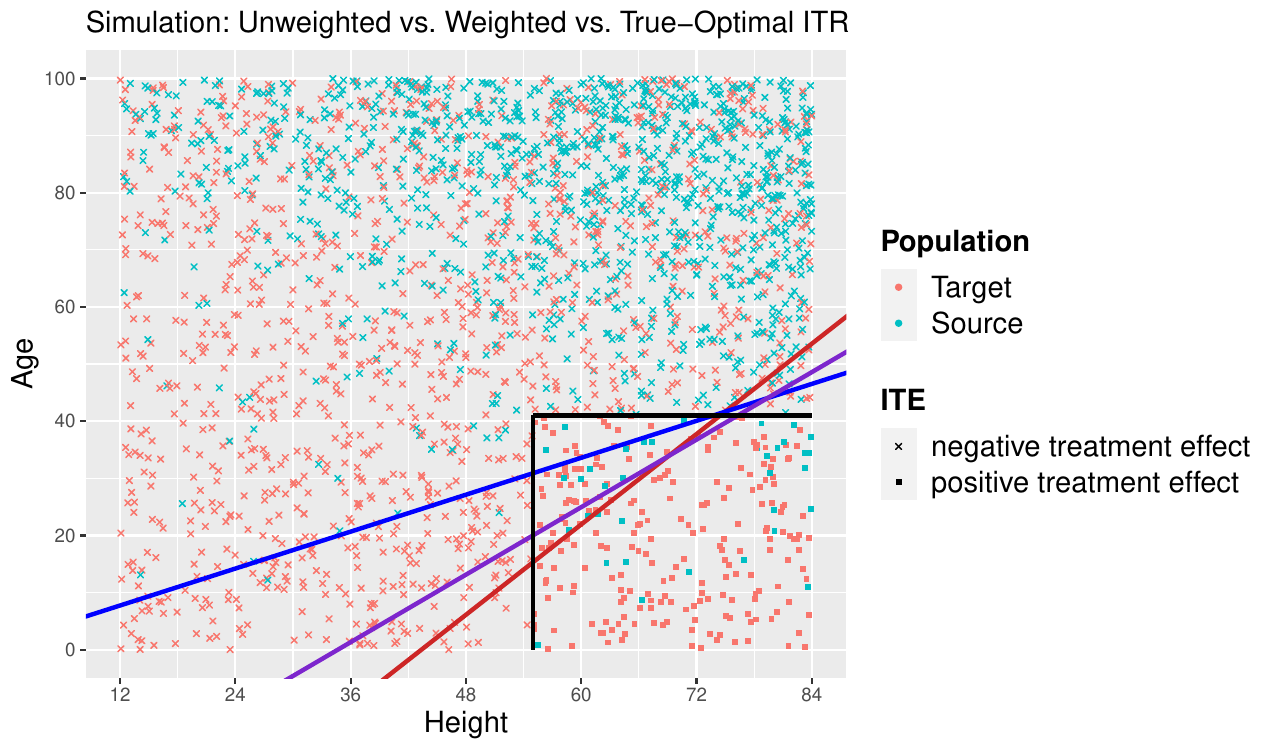}
    \caption{Calibration Weighting Improves Estimation of Treatment Effect (Blue Line: Unweighted ITR for the target population, i.e. True Optimal ITR for source population; Purple Line: Weighted ITR for target population; Red Line: True Optimal ITR for target population)}
    \label{fig:Figure3}
\end{figure}

In our simulation, we generate treatment and outcome information for the target population. This would not be available in a real scenario since our end goal is to assign the best treatment to patients in the target population to maximize their outcome. However, this information is useful to establish the best ITR for the target population as a standard to compare to. The true optimal linear ITR for the target population (shown in red in Figure \ref{fig:Figure3}) yields a 94.5\% correct classification rate, i.e. 94.5\% of the patients in the target population received the treatments that gave them the better outcome under the true optimal ITR. 

When we derive our ITR from only the source population information without considering the heterogeneity problem between the source and target populations, this ITR is the best ITR for the source population, but an unweighted ITR for the target population (shown in blue in Figure \ref{fig:Figure3}). Since heterogeneity of the covariate distribution remains, this ITR--although well fit for the source population--is not a good fit for the target population, having a correct classification rate of 86.0\%.

When we compute the calibration weights, we consider covariates, treatment, and outcome information on the source population, and covariates information on the target population, taking full advantage of the information we can obtain in a real-case scenario. The resulting calibration-weighted ITR (shown in purple in Figure \ref{fig:Figure3}) has an improved correct classification rate of 93.6\%. Compared to the unweighted ITR, it is much closer to the true optimal ITR for the target population.

\begin{figure}[h!]
    \centering
    \includegraphics[scale = 0.7]{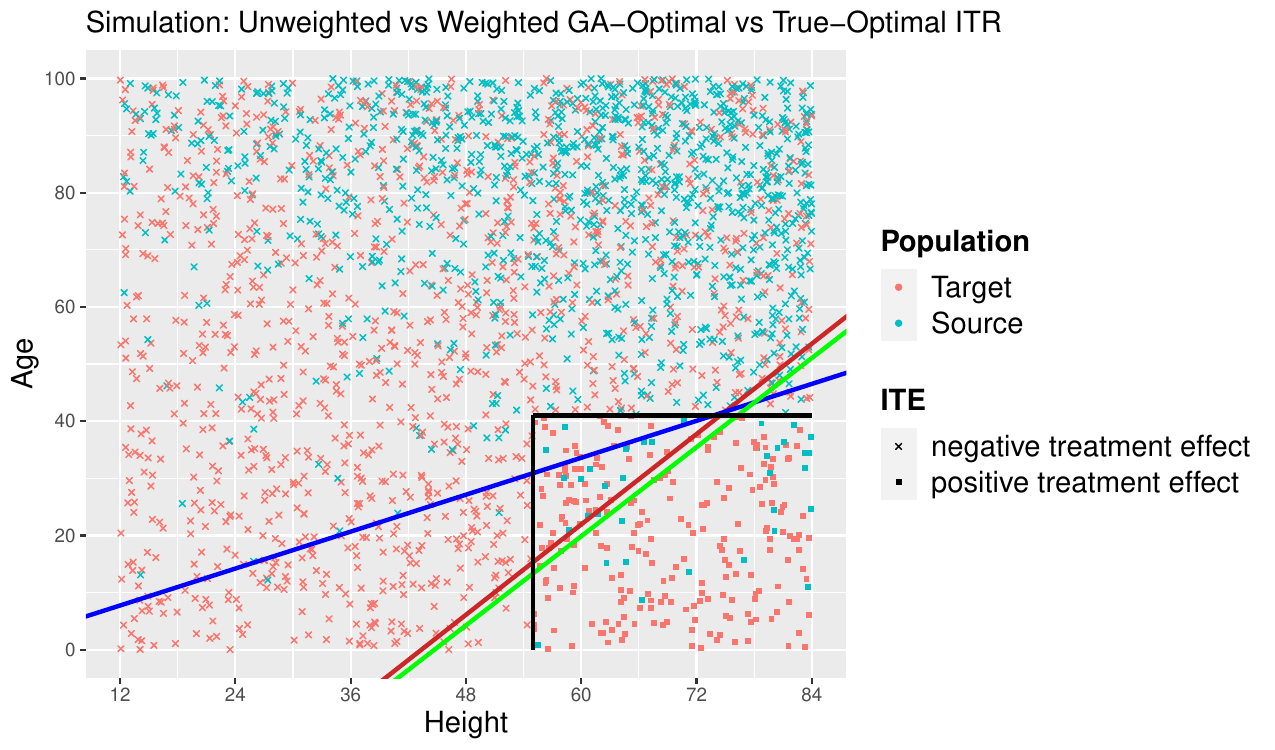}
    \caption{Combination Of Calibration Weighting And GA Optimization Yields A Close Approximation Of The True Optimal ITR (Blue Line: Unweighted ITR for target population, i.e. True Optimal ITR for source population; Green Line: Weighted ITR optimized with GA for target population; Red Line: True Optimal ITR for target population)}
    \label{fig:Figure4}
\end{figure}

After demonstrating the effect of calibration weighting, we moved on to the optimization problem to produce the optimal weighted linear ITR that maximizes patient outcomes. After 100 iterations of GA with 100 different random seeds that generate different samplings of the population pool, the algorithm yields an ITR with the largest value for the value function (shown in green in Figure \ref{fig:Figure4}), which closely approximates the true optimal ITR for the target population.

\section{Discussion}  \label{s:disc}

\subsection{Advantages} \label{ss:advn}

The transfer learning framework has many advantages and impacts for the statistical and clinical communities. When used appropriately, it can be a resourceful tool that bridges diverse sources and target populations, connecting data from various sources to account for patient heterogeneity, and informing more accurate decisions. It embraces the essence of precision medicine by providing precise treatment recommendations tailored to each patient's characteristics, assisting clinicians across health care systems globally to make data-driven treatment decisions for patients. In addition, transfer learning is a transferable framework and can be applied in other fields of study such as marketing, technology, social sciences, and education.

\subsection{Limitations} \label{ss:limi}

There are some limitations to the work we have done. A limitation in our proposed methods resides in the Genetic Algorithm (GA) we used to optimize the value function. Like many other optimization algorithms, GA can become computationally expensive when applied to large datasets. It is also sensitive to the choice of initial values, i.e. the randomly selected initial population of candidate solutions. Additionally, GA can only be applied under moderate covariate size. When there are too many covariates, the search space can be too large to find the global maximum. 

In our real-world application, we removed covariates from either dataset due to either discreteness or a high level (greater than 60\%) of missingness. The removed covariates include biomarkers such as arterial pH, PaO2 (the partial pressure of oxygen in the arterial blood), pCO2 (partial pressure of carbon dioxide), FiO2 (fraction of inspired oxygen), etc., which may provide useful information to assess whether mechanical ventilation should be used on individual patients to improve outcome. However, the large missingness of these biomarkers makes it difficult to perform data imputation. Therefore we had to discard these biomarkers from our model. Additionally, due to the lack of survival outcome information in the real-world data, we are unable to confirm the real-world efficacy of the treatment rule generated by our method.

\subsection{Future Work} \label{ss:futw}

In our proposed framework, we used logistic regression and linear regression for estimation of the propensity score and outcome regression components of CAIPW estimator. However, we plan to explore a variety of semi-parametric and nonparametric estimation techniques such as generalized additive models, spline regression, random forests, and Super Learners. These techniques may be more advantageous for capturing non-linear relationships, particularly within high-dimensional covariate-outcome spaces, further improving the robustness of the CAIPW estimator. 

Tree-based ITRs are a promising alternative to linear ITRs. In addition to maintaining interpretability, ITRs in a decision-tree form more closely mimics human decision-making, which will assist communication between statisticians and medical professionals. A tree-based rule also makes it straightforward to select important variables and reduce the rule size through pruning.

\section*{Supplementary Material} \label{s:supp}

Supplementary material includes a README file that instructs users how to use the supplementary material, pre-processed eICU-CRD and MIMIC-III data files for the medical application, an R script containing the R functions, and R markdown files for both the simulation study and the medical application.

\section*{Acknowledgments}  \label{s:ackn}
The authors gratefully acknowledge the generous support from National Science Foundation (NSF) grant DMS2051010 and National Security Agency (NSA) grant H98230-22-1-0006. This research is also supported in part by the National Institute of Environmental Health Sciences (NIEHS) training grant T32ES007018.

\newpage

\bibliographystyle{apalike}
\spacingset{1}
\bibliography{TLITR}

\end{document}